\documentclass[a4paper,12pt]{article}
\usepackage[cp1251]{inputenc}
\usepackage{graphics}
\usepackage{amssymb,amsmath}
\usepackage[english]{babel}
\usepackage{indentfirst}
\begin{document}
\bigskip
\centerline {\bf On the global magnetic activity and dynamo of the
Sun and solar-type stars}

\bigskip

\centerline {E.A. Bruevich $^{a}$ , I.K. Rozgacheva $^{b}$}
\bigskip

\centerline  {$^a${Moscow State University, Sternberg Astronomical
Institute, Russia}}
\centerline {$^b${Moscow State Pedagogical
University, Russia}}
\centerline {E-mail: $^a${red-field@yandex.ru},
$^b${rozgacheva@yandex.ru}}

\bigskip
{\bf Abstract.}
      The activity of the Sun as a result of cyclic changes of the global magnetic field is studied.
      As a consequence of the analysis of magnetic activity of solar-type stars the following power dependencies were found:
      the dependence between the rotation periods and the effective temperatures $P_{rot} \sim T_{eff}~^{-3,9}$,
      the dependence  between the duration of the "11-year" cycles of activity and the effective temperatures  $~T_{11} \sim T_{eff}~^{-1,1}$
      and the dependence  between the duration of quasi-biennial cycles and the effective temperatures $T_{2} \sim T_{eff}~^{-0,79}$.
      It is shown that the physical nature of these  dependencies associated with the observed properties of solar-type stars
      and can be explained by the existence of internal Rossby waves around the base of convective shells of these stars.

\bigskip

\bigskip

\vskip12pt {\bf1  The magnetic activity of the Sun }
\vskip12pt
Magnetic activity of the Sun is called the complex of
electromagnetic and hydrodynamic processes in the solar atmosphere.
The analysis of active regions (plages and spots in the photosphere,
flocculae in the chromosphere and prominences in the corona of the
Sun) is required to study the magnetic field of the Sun and the
physics of magnetic activity. This task is of fundamental importance
for astrophysics of the Sun and the stars. Its applied meaning is
connected with the influence of solar active processes on the
Earth's magnetic field.

It is difficult to predict the details of the evolution of each
active region in present time. However, the total change in active
regions is cyclical. Long known "11-year" cycle of solar activity,
the duration of which varies from 7 to 17 years. Quasi-biennial
cycles of solar and solar-type stars magnetic activity were found
(Vitinsky et al. 1986; Rivin 1989; Khramova et al. 2002; Rozgacheva
and Bruevich 2002; Kolah \& Olach 2009; Bruevich \& Kononovich 2011;
Bruevich \& Ivanov-Kholodnyj 2011; Bruevich \& Rozgacheva 2012).

 In the theory of the solar dynamo (Parker
1979; Monin 1980; Vanshtein et al. 1980; Makarov 1987; Berdugina et
al. 2006) (Babcok - Layton  $\alpha\Omega$- dynamo model), the
magnetic activity of the Sun is explained with the help of two main
effects: first - production of azimuth (toroidal) field with help of
a large-scale field (poloidal) due to the differential rotation of
the convective envelope ($\Omega$ - effect); secondly - formation of
the  poloidal field with help of  some local bipolar magnetic
regions of the toroidal field ($\alpha $ - effect) due to
differential rotation.

It is supposed, that in the maximum of the "11-year" cycle the old
poloidal field, from which it was generated, a toroidal field, has
already disappeared, and the generation of new poloidal field
begins. In the theory of the solar dynamo the hypothesis of a strong
turbulent convection in the under-photospheric layers is used.
Strong turbulent convection is necessary requirement for the
effective generation of magnetic fields. The theory of the solar
$\alpha~\Omega$ - dynamo well simulates the following phenomena of
local magnetic activity on the Sun and the stars:

$\bullet$   the formation of strong local magnetic fields (of the
order of 0,1 Tesla), the evolution of which leads to the appearance
and evolution of spots and plages in photosphere and the energetic
events in the chromosphere and corona, such as prominences, flares
and coronal mass ejections;

$\bullet$   the cyclicity of magnetic activity (on the basis of the
selection of the parameters of the turbulent viscosity tensor);

$\bullet$   the Shperer's law ("active latitudes" of spot's
appearance during an "11-year" cycle  move to the sun's equator,
from helio-latitudes equal to   $\pm52^{\circ}$ for beginners
cycle's spots to helio-latitudes equal to   $ \pm5^{\circ}$ for the
final cycle's spots);

$\bullet$  "11-year" cycles are closely connected with each other,
and partially overlap, because near the minima of the activity, when
the last spots of the old cycle are still visible at the equator of
the Sun, the spots of a new cycle has already appeared on the high
helio-latitudes;

$\bullet$   at the graph of dependence of the annual Wolf numbers on
time, the duration of the branch growth cycle decreases with
increasing amplitude of the cycle, and the area under the branch of
the decline of the cycle increases with the amplitude of the cycle.

To the present time the following facts on the global magnetic
activity of the Sun, which covers the whole of the outer layers of
the Sun and has a complex evolution over time are known:

$\bullet$  active processes of the transformation of magnetic energy
(plages, flocculi, flares)  happen always and in all the outer
layers of the Sun (Obrydko 1985; Obrydko 1999);

$\bullet$ there exist the "active longitudes", near to which there
is high concentration of the spots. They are quasi-periodically
distributed on the helio-meridians and over time, slowly moving
during of the "11-year" cycle (Obrydko et al. 2009; Makarov et. al
2001; Schnerr \& Spruit 2010); Also the "active longitudes" slowly
movements during the 11-year cycle were detected in UV and X-rays by
SOHO satellite observations (SOHO 1995-2012).

$\bullet$ duration of the "11-year" cycles are changed, and, the
larger the amplitude of the cycle, the shorter the cycle;

$\bullet$  with the increasing of the amplitude of the "11-year"
cycle the starting "active latitudes" will be increased  (up to
$\pm60^{\circ}$);

$\bullet$ there is the "Gnevyshev-Ol's rule"  - in the 22-year
magnetic cycle the amplitude of the initial 11-year cycle (even
cycle) is always smaller than the amplitude of the final 11-year
cycle (odd cycle);

$\bullet$  active cycle with a duration of 1,3 years on the
background of the last eight of the "11-year" cycles was detected
(Livshits \& Obrydko 2006)

$\bullet$  analysis of cyclic variations of the chromospheric
radiation of solar-type stars allows us to detect a set of cycles
with duration from 2 years to 15 years (2-3 years, 5-6 years, 8-13
years) (Bruevich \& Ivanov-Kholodnyj 2011);

$\bullet$  series of Wolf numbers observations for the majority of
the cycles have two almost identical maximums;

$\bullet$ the Sun's flare activity is an important indicator of the
general level of activity of the atmosphere is also described in
other activity indices, in particular around the solar disk index
and a locally varying flux in the H-alpha, see Bruevich (1995).

$\bullet$ coronal holes, polar plages at the latitudes of  $\vert
\varphi \vert > 40^{\circ}$  which are characterized  by  strong
magnetic fields (the intensity of the magnetic fields is of the
order to the kilogauss) are observed as features of  high-latitude
and polar activity;

$\bullet$  in the unperturbed chromosphere we can always observe
spicules and supergranulation, number of spicules is the most of all
at the poles (on 30\% more, than at the equator) and least of all in
the latitude of $35^{\circ}$ (on 10\% less than at the equator);

$\bullet$  the polar spicules are inclined to the equator, in the
"active latitudes" spicules are inclined to the nearest pole (follow
the direction of the magnetic field);

$\bullet$ the dipole component of the large-scale magnetic field
during the maximum of local activity  was detected. The axis of the
dipole was located in the equatorial plane of the Sun;

$\bullet$  during the minima of the local activity the regions of
opposite polarity of the magnetic field   are observed. They
alternate on longitude with prevailing wave number $m = 6$ (the
giant convection cells).

The above facts on the global magnetic activity indicate that, in
the convective shell of the Sun work not less than two mechanisms
$\alpha~\Omega$ - dynamo. They have a variety of spatial scales and
different characteristic times. These mechanisms should ensure the
generation of several magnetic fields of different scales - the
strong local magnetic fields, and weaker (to three orders of
magnitude) large-scale global field.

In the framework of the hypothesis on one turbulent convective shell
it is impossible to ensure sustainable separation of small-scale and
large-scale hydro-magnetic dynamos. This hypothesis does not allow
to explain the ordering of processes of activity in time (cycles)
and in space ("active latitudes and longitudes", granulation and
supergranulation). For example, Kolmogorov's spectrum of developed
turbulence is not implemented for speed and size of granules
(Nordlund et al. 1997). This requires the further development of the
theory of $\alpha~\Omega$ - dynamo (Spruit 1998; Spruit 2004).

At the present time the observational tests that can confirm or
clarify the main hypothesis of the theory of the solar dynamo are
necessary. So far the basic test is considered to be the existence
of magnetic activity of the Sun.

A statistical analysis of the data on of the color indices, periods
of rotation, the duration of activity cycles, the magnetic field,
the age of solar-type stars is necessary. Such research will help to
find a relation between the parameters of cycles such as the
duration of activity cycles, the amplitude of the variations of the
fluxes of radiation of different indices of activity   with the
physical parameters of the stars.

At present the observations of active atmospheres of solar-type
stars are regularly held in the framework of the "HK-project" at
Mount Wilson observatory (Baliunas et al. 1995; Radick et al. 1998;
Lockwood et al. 2007)

In the work (Bruevich \& Rozgacheva 2012) we published the results
of processing of the observations of cyclic variations of
chromospheric radiation of the 52 solar-type stars and the Sun  in
the form of a Table. In this Table an information about the periods
of rotation and effective temperatures of stars, about the duration
of their "11-year" and quasibiennial cycles is contained. The
periods of cycles of activity are given according to the
calculations of the authors of this work together with the
definition of "11-year" periods with help the authors of
"HK-project" (Baliunas et al. 1995). We have to point out that close
interconnection between radiation fluxes characterized the energy
release from different atmosphere's layers is the widespread
phenomenon among the stars of late-type spectral classes. Bruevich
\& Alekseev (2007) confirmed that there exists the close
interconnection between photospheric and coronal fluxes variations
for solar-type stars of F, G, K and M spectral classes with widely
varying activity of their atmospheres. It was shown that the sum of
areas of spots and the values of X-ray fluxes increase gradually
from the Sun and HK project stars with the low spotted discs to the
highly spotted K and M-stars. The variations of activity indices in
the whole 11-yr cycle of the Sun are very similar to the cyclical
variations of the chromospheric fluxes on the stars. So we can
simulate the dependencies which describe the variations of the
indices during the activity cycle for the stars as for the Sun, see
Bruevich and Bruevich (2004).

In the present work for a sample of 52 stars from (Bruevich \&
Rozgacheva 2012) we found the statistically significant dependencies
"the rotation period - the effective temperature" and "the duration
of activity cycle - the effective temperature" (see below, section
2).

Next, in section 3 we give a physical interpretation of these
dependencies on the basis of the hypothesis about the possibility of
the existence of several layers in the convective shells of the
solar-type stars.

Our physical interpretation is based on the assumption of the
existence of not less than two  layers located  under the
photosphere.

At the bottom level of convective zone the layer of the laminar
convection is situated. This layer    consists of giant convective
cells. Due to differential rotation the Rossby waves with non-zero
helicity are formed on the surface of this layer.

The formation of the laminar convection is a consequence of a strong
heating of plasma with help of photons from the zone of radiant heat
transfer and of the high radiant viscosity of fully ionized dense
plasma. Radiant viscosity inhibits the directed motion of electrons
more effectively than directed motion of protons. Therefore, the
current appears in the Rossby waves, and this current generates the
poloidal field. This field is a primary field for the whole of the
magnetic activity of stars.

 Above the layer of the laminar convection the layer of turbulent convention is situated. In the layer of turbulent
 convection primary poloidal field generates a toroidal field and starts the mechanism of generation
 of a strong local magnetic activity at the medium and equatorial latitudes of solar-type stars.
 Local and global magnetic activity are interconnected thanks to the existence of internal
 Rossby waves and thanks to the primary poloidal field.

\vskip12pt {\bf2  Dependencies "the rotation period - the effective
temperature" and "the duration of the cycle of activity - the
effective temperature" } \vskip12pt

We used the data of observations of variations of chromospheric
radiation of the Sun and of 52 solar-type "HK-project" stars from
the Table, published in (Bruevich \& Rozgacheva 2012) for
statistical analysis and the search if there is a possible linear
relationship between the periods of rotation Prot of the stars and
their effective temperature $T_{eff}$. Diagram "the rotation period
- the effective temperature" for 52 solar-type stars is shown in
Fig. 1.

\begin{figure}[tbh!]
\centerline{
\includegraphics{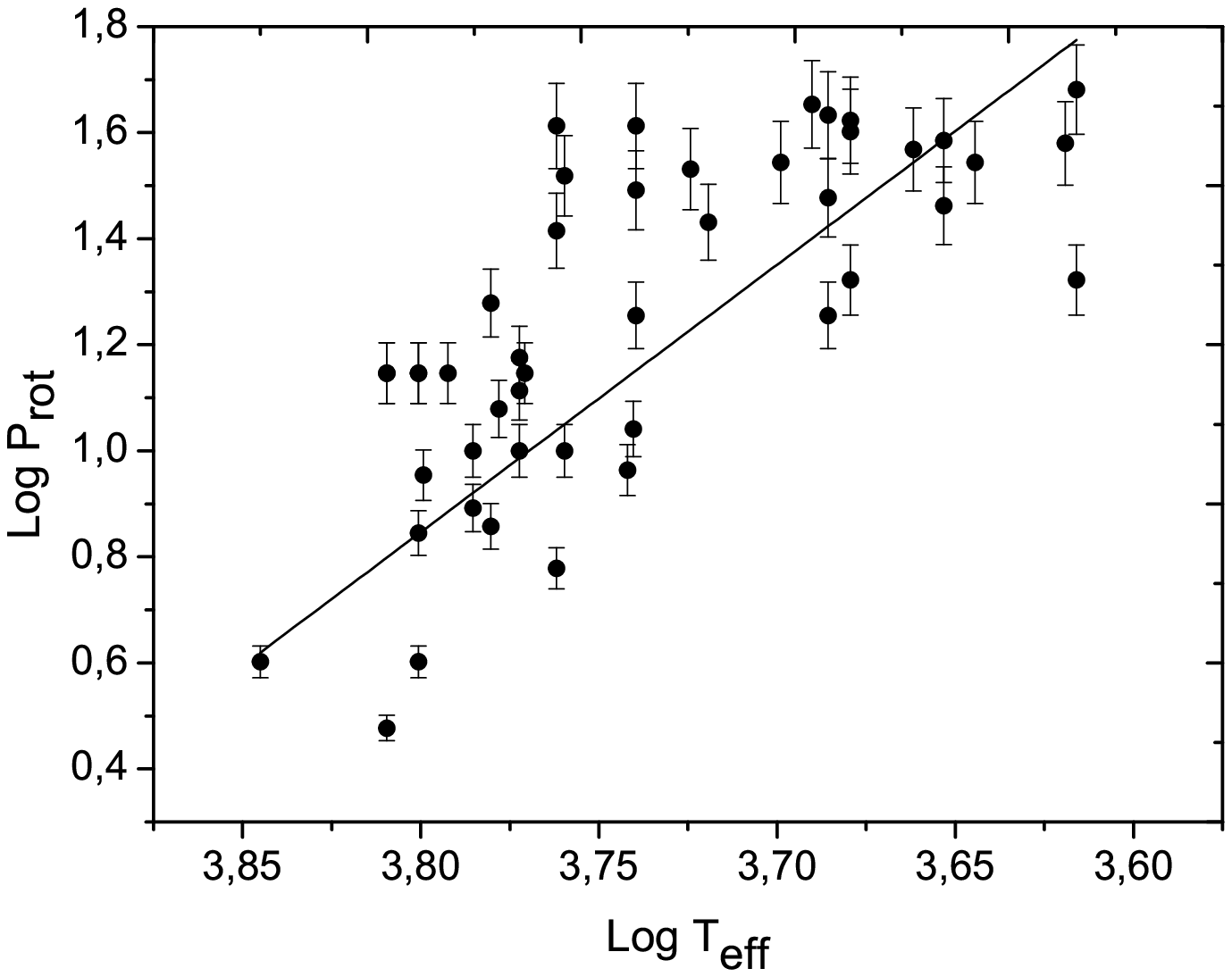}}
 \caption{Diagram "the rotation period - the effective temperature".
The line shows the linear regression on a data set}
{\label{Fi:Fig1}}
\end{figure}

Linear regression equation has the following form:
$$ \log P_{rot}=15,7 -  3,87 \cdot \log T_{eff}   \eqno (1)$$

The linear correlation coefficient (Pearson's correlation
coefficient) in the regression equation (1) is equal to 0,73.
According to Pearson's cumulative statistic test (which
asymptotically approaches a $ \chi^2$ -distribution) the linear
correlation between the $P_{rot}$ and $T_{eff}$  is statistically
significant at a 0,05 level of significance.

Thus our data set of the rotation periods and the effective
temperatures of stars shows the    following power-law dependence
$$ P_{rot} \sim T_{eff}~^{-3,9}  \eqno (2) $$

The diagram of  "the duration of the cycle - the effective
temperature" ($T_{11}$ is the period of   " 11-year" cycles) for 46
stars from Table, published in (Bruevich \& Rozgacheva 2012) is
shown in Fig. 2.

\begin{figure}[h!]
\centerline{
\includegraphics{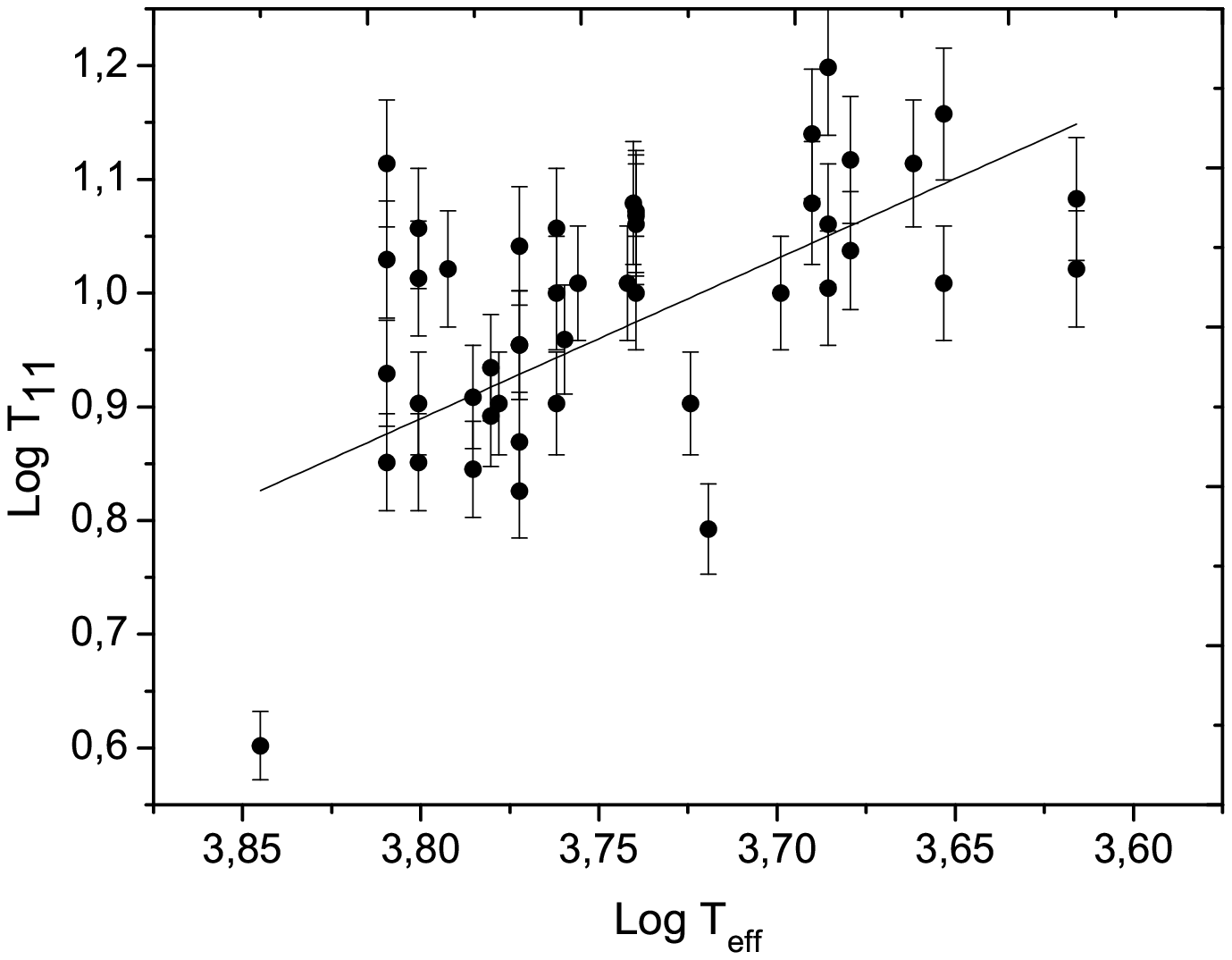}}
\caption{Diagram "the duration of the cycle $T_{11}$ - the effective
temperature $T_{eff}$"  diagram. The line shows the linear
regression on a data set }\label{Fi:Fig2}
\end{figure}

The linear regression equation for points of diagram at the Fig. 2
is of the form:

$$ \log T_{11}= 5,15 -  1,11 \cdot \log T_{eff}   \eqno (3)$$

The linear correlation coefficient (Pearson's correlation
coefficient) in the regression equation (3) is equal to (- 0,67).
According to Pearson's cumulative statistic test (which
asymptotically approaches a
 $ \chi^2$ -distribution)
the linear correlation between the $T_{11}$  and $T_{eff}$ is
statistically significant at a 0,05 level of significance.

Thus, for the investigated sample of stars the periods of  "11-year"
cycles $T_{11}$   and their effective temperatures $T_{eff}$  are
connected in power-law dependence:
$$ T_{11} \sim T_{eff}~^{-1,1}  \eqno (4) $$

"The duration of the cycle - the effective temperature"  diagram
($T_2$ is the period of   quasi-biennial cycles) for 27 stars from
Table, published in (Bruevich \& Rozgacheva 2012) is shown in Fig.
3.

\begin{figure}[h!]
\centerline{
\includegraphics{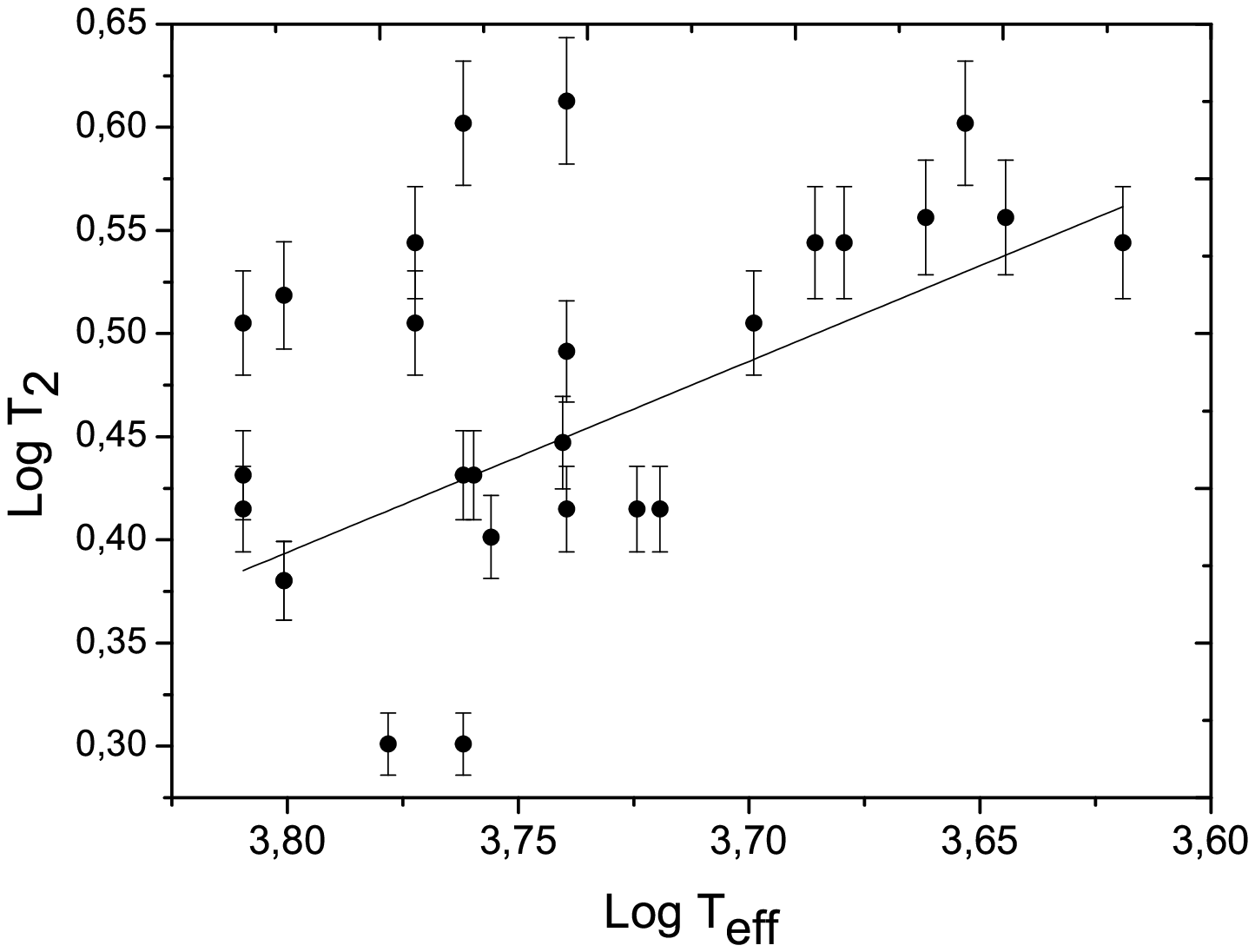}}
\caption{Diagram "the duration of the cycle $T_{2}$ - the effective
temperature $T_{eff}$"  diagram. The line shows the linear
regression on a data set }\label{Fi:Fig3}
\end{figure}

Linear regression equation for points of diagram at the Fig. 3 is of
the form:
$$ \log T_2 = 3,46 -  0,79 \cdot \log T_{eff}   \eqno (5)$$

The linear correlation coefficient (Pearson's correlation
coefficient) in the regression equation (4) is equal to  (- 0,51).
According to Pearson's cumulative statistic test (which
asymptotically approaches a $ \chi^2$ -distribution ) the linear
correlation between the $T_2$  and $T_{eff}$  is statistically
significant at a 0,1 level of significance.

Thus, for the investigated sample of stars their periods of
quasi-biennial cycles $T_2$ and their effective temperatures
$T_{eff}$ are linked as power-law dependence:

$$ T_2 \sim T_{eff}~^{-0,79}  \eqno (6) $$

\vskip12pt {\bf3  Physical model of "the rotation period - the
effective temperature" dependence}
\vskip12pt

The connection between the rotation period and the effective
temperature of the stars comes from the following facts:

Firstly, the continuous spectrum of radiation of the Main sequence
stars at the Hertzsprung-Russell diagram is well approximated by the
Plank's formula for the radiating plasma, which is in thermal
equilibrium with its own radiation. Therefore, we use the Stefan -
Boltzmann law for the luminance of the surface of the stars:

$$ \frac{L_{star}}{L_{Sun}}=\left( \frac{R_{star}}{R_{Sun}} \right)^2
\left( \frac{T_{eff}~^{star}}{T_{eff}~^{Sun}} \right)^4    \eqno (7)
$$

where $L_{star}$, $L_{Sun}$ - are the luminosities of stars and the
Sun, $R_{star}$, $R_{Sun}$ are the radii of the stars and the Sun,
$T_{eff}~^{Sun}$ - is the effective temperature of the Sun,
$T_{eff}~^{star}$ - are the effective temperatures of the stars.

Secondly, for Main sequence of stars there is dependence of the
"mass of the star - luminosity of the star", which is well
approximated by a power-law dependence:

$$ L \sim M_{star}~^{\alpha}  \eqno (8) $$

Where $M_{star}$ is the mass of the star, the exponent $\alpha$
depends on the mass of the star and its spectral class, for stars
with masses similar to the mass of the Sun  $ 3,5 \leqslant \alpha
\leqslant 4 $.

In the third place, in the course of the evolution of the single
stars of the Main sequence its angular momentum slowly decreasing
due to the loss of weight, because of interactions with
interplanetary plasma and to the planets. A significant decrease of
the angular momentum comes slowly for billions of years.

In a relatively short time intervals, which are characteristic of
magnetic activity, we can assume that the angular momentum is
maintained.

In this case $M\cdot R_{star}\cdot \frac{2\pi R_{star}}{P_{rot}}\sim
const$ where $P_{rot}$ is the rotation period. We assume for
simplicity $P_{rot}$ of the Sun is $P_{Sun}$, $P_{rot}$ of the star
is $P_{star}$.

Therefore, we can write the relationship between mass, radius and
period of rotation:

$$ \frac{M_{star}}{M_{Sun}}=\left(\frac{R_{star}}{ R_{Sun}}\right)^{-2}
\left(\frac{P_{star}}{ P_{Sun}}\right)    \eqno (9) $$

From the formulae (7) - (9) follows that there is the relationship
of the period of rotation and the effective temperature:

$$ \frac{P_{star}}{1  year}=\left(\frac{T_{eff}~^{star}}{ 1K}\right)^{-4}
\left(\frac{M_{star}}{ M_{Sun}}\right) ^{\alpha+1}   \eqno (10) $$

Thus, at the diagram "the period of rotation of - the effective
temperature" the solar-type stars should be placed near the line
with the equation:
$$ \log\left(\frac{P_{star}}{1  year}\right) = - 4 \log \left(\frac{T_{eff}~^{star}}{ 1K}\right) + const   $$

So the above regression equation (1) agrees well within the errors
with the theoretical formula (10).

\vskip12pt {\bf4  Physical model of the dependency "the duration of
the cycles of the stars - the effective temperature of the star"}
\vskip12pt

In the work (Seehafer et al. 2003) a numerical MHD simulation of
hydromagnetic dynamo was made. The authors have studied the fully
turbulent and the fully laminar convective shells. They also studied
the convective shell which consists of the shell of two layers: the
turbulent convection and the laminar convection. Found that the
numerical model of the solar dynamo is consistent with the
observations of local magnetic activity, if in the convective
envelope there are the layers both of the laminar and turbulent
convection. Therefore it is likely that the convective shell of the
Sun can be stratified into layers with different types of
convection. We use this model result for the description of a
physical nature of hydromagnetic dynamo of different scales.

The above properties of the global magnetic activity of the Sun and
the power dependencies between the duration of cycles and effective
temperatures of solar-type stars can be explained under the
following scheme of physical processes in their convective shells:

\begin{figure}[h!]
\centerline{
\includegraphics{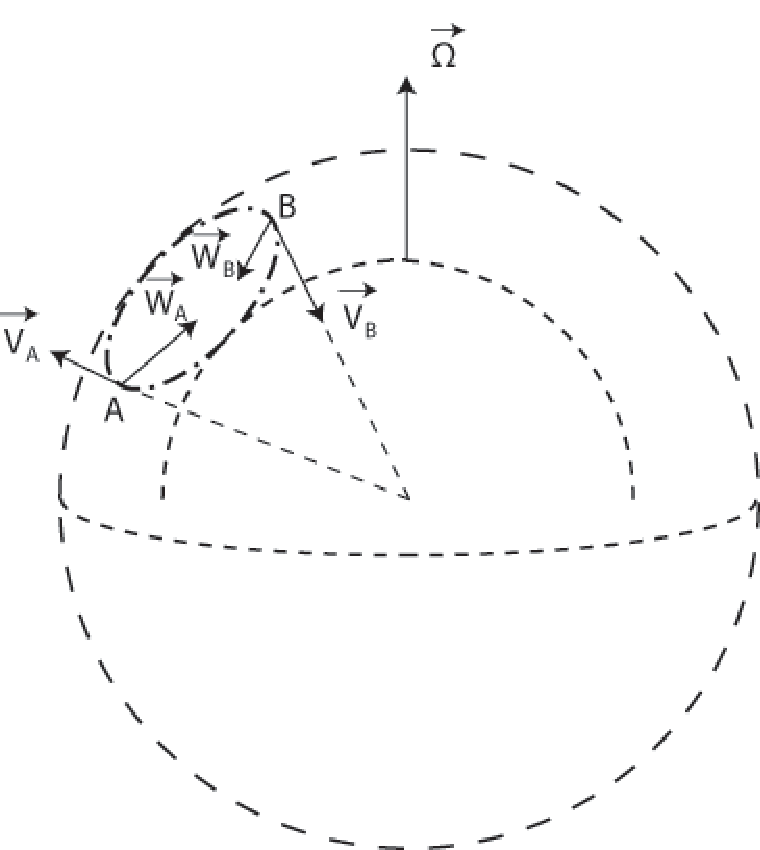}}
\caption{Scheme of the directions of the vectors of velocities
($\vec{V_A}$ and $\vec{V_B}$) and the vectors of the Coriolis
accelerations ($\vec{W_A}$ and $\vec{W_B}$) in the convective cell
}\label{Fi:Fig4}
\end{figure}

(1)  The process of magnetic activity begins with the formation of
giant convective cells in the layer near the base of the convective
envelope due to the heating of the plasma with help of photons
coming from the radiation zone. In the work (Bruevich \& Rozgacheva
2010) it is shown that near the base of convective envelope of the
Sun the conditions for convective transfer from the account of the
radiant viscosity of plasma only are fulfilled.

When the plasma elementary volume is coming to light due to
convection along the radial direction its path is rejected from the
radial direction by the Coriolis forces. Figure 4. shows the
directions of the Coriolis acceleration $\vec{W_A}$ of rising
elementary volume of plasma  at the point A and the Coriolis
acceleration $\vec{W_B}$   of falling elementary volume at the point
in the in the convective cell.  Rising element of plasma in the
point A is located closer to the equator than falling element of
plasma at the point in the point B. The Coriolis accelerations are
directed perpendicular to the radial plane, in which moves an
elementary convective volume. The Coriolis acceleration in the point
A is more than Coriolis acceleration at the point B, because the
angular velocity of rotation increases with decreasing of latitude.

The rising element of volume creates a pressure gradient, the
direction of this gradient  depends on the direction of the velocity
of the element and on the Coriolis acceleration, on the change of
speed of rotation with the latitude (increasing of the speed of
rotation in the direction of the equator or at the direction of the
poles). The pressure gradient will be spread in the spherical shell
along the lines of latitudes as Rossby wave.

The first model with the use of Rossby waves in the theory of
hydromagnetic dynamo was proposed in work (Gilman 1974). He
considered that the Rossby waves were elements of spiral turbulence
in the rotating convective shell, which occurs due to the
latitudinal temperature gradient. Later this hypothesis was refined.
In the total it was suggested to consider the Rossby waves,
resulting from the vertical temperature gradient (Monin 1980).

In the work (Schmitt 1987) the solution of model equations for
Rossby waves was found, thanks to which a toroidal field can be
generated by a poloidal field.

Around the base of convective shell the plasma temperature reaches
millions of degrees $T\approx 2\cdot 10^6~K$.  Plasma is fully
ionized at such temperatures. Interaction of radiation with plasma
is carried out by the scattering of photons by electrons; if the
characteristic energy of photons does not exceed $kT$ (where k is
the Boltzmann constant). The length of free photon scattering order
$\left(\sigma_T n \right)^{-1} \approx 3\cdot 10^2 ~cm$ when
concentration of plasma is equal  about  $n\approx 5\cdot 10^{21}
cm^{-3}  $, where $\sigma_T $ is the cross-section of Thomson
scattering of photons by electrons.

The motion of the plasma in the Rossby wave is slowed by the radiant
viscosity of plasma. This  viscosity inhibits the directed motion of
electrons (the characteristic time is a fraction of a second),
faster than motion of protons. Therefore, in the Rossby wave the
electric current appears. This current creates a poloidal field.
Lines of force of this poloidal field have the wave-like structure.
This field is concentrated around the base of convective envelope in
the equatorial and medium-sized helio-latitudes. In the polar
helio-latitudes lines of force of the poloidal field come to the
atmosphere of the Sun (Fig. 5).

\begin{figure}[h!]
\centerline{
\includegraphics{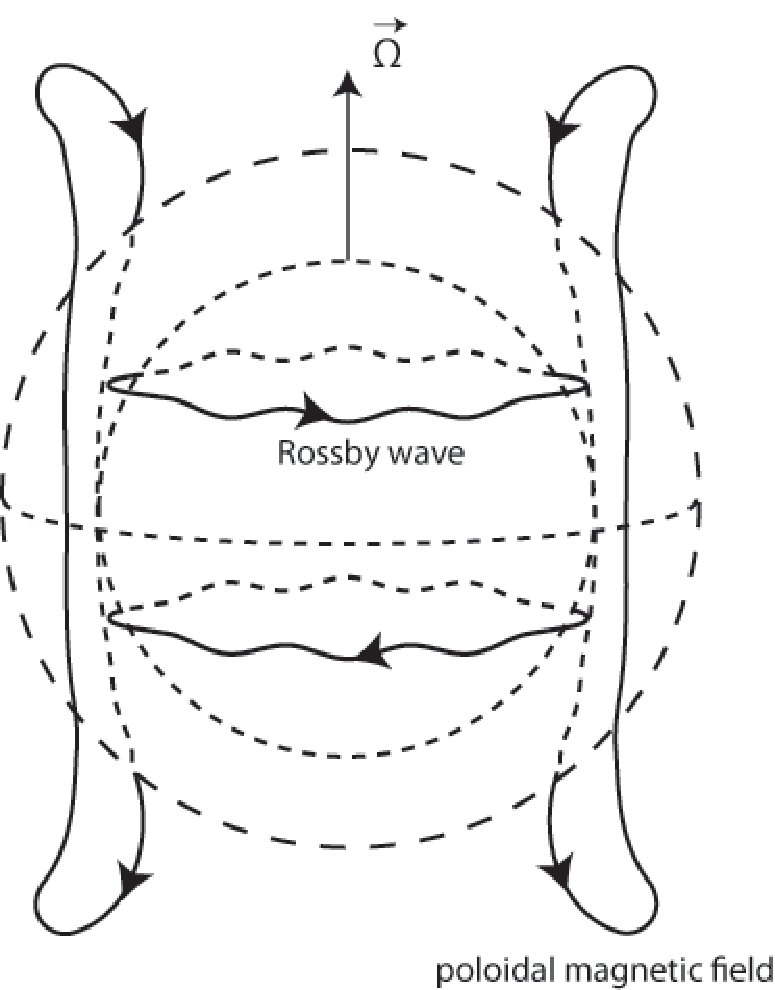}}
\caption{The scheme of Rossby waves and poloidal field
}\label{Fi:Fig5}
\end{figure}

On the length of the latitude the whole number of Rossby waves is
packed. Therefore, the length of these waves should be of the order
of  $\lambda \approx (1/m) 2 \pi R_I \cos \phi $  where $\phi$ is
the latitude of parallels, along which the wave applies, $R_I\approx
0,7 R_{Sun}$   is the radius of the base of the convective envelope,
m is the integer number of  Rossby waves. For giant cells which are
observed in photosphere  the wave number is of order $m=6$ (Monin
1980).

The wave length depends on the latitude. At different latitudes the
various Rossby waves are generated, in each of which the electric
current appears. These currents generate the poloidal magnetic
fields. The ordered geometric structure of the spicules which are
observed in the high helio-latitudes, points to the regularity of
the poloidal field. Therefore, the lines of force of poloidal
fields, created by Rossby waves, must be regular and not be
entangled due to the large-scale convection.

It will be in that case, if the convective cells are big enough and
are about the same, i.e. the convection around the base of
convective envelope must be laminar. The length of the Rossby wave
is approximately equal to the thickness of the shell of laminar
convection.

We estimate the characteristic time of formation of the Rossby
waves. The characteristic time of the convective ascent of the
plasma element which is heated by photons from the radiation zone is
equal:

$$ t \approx \frac {\nu}{gh \left( \frac {1}{\rho} \frac {\partial \rho}{\partial T} \right) \Delta
T} \sim const $$

where $\nu$  is the coefficient of viscosity of the plasma, $g$ is
the free fall acceleration, $h$ is the thickness of the shell of
laminar convection,  $ \left( \frac {1}{\rho} \frac {\partial
\rho}{\partial T} \right) $  is coefficient of thermal expansion of
the plasma, $ \Delta T$  is the gradient in the layer.

The average Archimedean acceleration of a plasma element is
approximately equal to $$ a\approx g  \left( \frac {1}{\rho} \frac
{\partial \rho}{\partial T} \right) \Delta T $$. Then the average
Archimedean speed of a plasma element is $ V \approx at \approx
\frac{\nu}{h}$.

The Coriolis forces turn and stretch the convective cell along a
parallel, so that the length of the path of element of the plasma is
comparable with the length of parallels $l=2 \pi \cos \varphi$.

Average acceleration of the Coriolis force on this Parallels is in
the order of magnitude is equal to $a_c= 2V\Omega_{Sun}\sin \varphi
\approx 2 \frac {\nu}{h}\Omega_{Sun}\sin \varphi$  , where $
\Omega_{Sun}$  is the angular velocity of rotation of the Sun around
the base of convective shell.

The time interval which is necessary for the generation of Rossby
waves ($t_R$) and poloidal magnetic field ($t_g$) on the order of
magnitude is equal to:
$$t_g \approx \sqrt{\frac{2l}{a_c}} \approx
\sqrt { \frac { \pi R_I \cos \varphi}{\frac{\nu}{h}\Omega_{Sun} \sin \varphi}}  \eqno (11) $$

For different latitudes the time of magnetic field formation is
different: the less latitude, the longer the time of the formation.
Therefore, the primary poloidal magnetic field should appear at high
latitudes. It can stimulate the polar magnetic activity.

(2) In accordance with the hypothesis of the Babcock - Layton
(Kichadinov 2005), we consider that the large-scale poloidal field,
connected with Rossby waves have to be the primary for the whole
complex of phenomena of magnetic activity.

From this poloidal field a toroidal field is generated due to the
differential rotation of the convective zone. The direction of the
magnetic force lines is the same as the direction of Rossby waves.
It is formed in the middle turbulent layer of the convective zone.
Here an  $\alpha \Omega$ - dynamo model works. Magnetic tubes and
loops of toroidal fields are generated as a consequence of turbulent
convection and also due to the curves of the magnetic force lines of
the primary poloidal field. Magnetic tube and loops of toroidal
fields rise to the photosphere. Their intersection and reconnection
stimulate the magnetic activity (Somov 2006).

 Rossby wave interacts
with the local magnetic fields which float due to the convection up
to the photosphere. This interaction compresses the region of
localization of the fields, increases the intensity of magnetic
fields and stimulates the formation of spots (Bissengaliev et al.
2010). So Rossby waves lead to the formation of "active latitudes"
(Monin 1980). Offset of local fields with help of the waves leads to
the formation of "active longitudes" with almost periodic
distribution along the helio-meridians.

The meridional fluxes around the base of convective zone are
directed from the poles to the equator (Kichatinov 2005). They move
Rossby waves closer to the equator. The consequence of this will be
the slope of the lines of force of poloidal field in the direction
to the equator. Generally speaking, the shifts of the currents which
are in the Rossby waves change the orientation of poloidal fields in
space: the axis of symmetry of the poloidal field turns and may be
situated in one of the surfaces of the equatorial latitudes.

The meridional fluxes around the base of convective zone are
transmitted to magnetic tubes of toroidal field due to the magnetic
viscosity of plasma. The consequence of this is observed in the
photosphere the "Shpherer's law "  which describes the local
magnetic activity on low helio-latitudes (each new cycle begins at
high latitudes in both hemispheres, gradually shifting to the
equator as the cycle unfolds).

In both hemispheres of the Sun photospheric bipolar magnetic regions
of toroidal fields are stretched, reconnected and form the poloidal
field of opposite polarity in relation to the primary poloidal
field. Then a toroidal field disappears. The first half of the cycle
of magnetic activity ends.

(3) New poloidal field interacts with the primary poloidal field.
Because these fields have the opposite direction, so the
reconnection of lines of magnetic takes place. At this time, there
appeared a magnetic activity above the middle latitudes. As a result
of this the processes of activity there are all over the surface of
the Sun.

Because of the rotations of the primary poloidal field a large-scale
field, which is formed by adding of two poloidal fields, may have a
complex structure, for example, quadruple structure.

In the general case a poloidal field is not completely destroyed due
to the polar activity. However, the new poloidal field becomes
weaker due to the polar activity.

The new toroidal field, which appears as a consequence of $\alpha
\Omega$ - dynamo from
 weakened poloidal field, will be weaker than the previous
 one toroidal field, which appeared from the primary poloidal field.
 Therefore, the local magnetic activity, which will be caused by this
 toroidal field in the second half of the magnetic cycle, will also be weaker.

 This is a new a toroidal field which is directed opposite to the direction of
 Rossby wave propagation. It weakens the electric currents that exist in these waves.
 This leads to a weakening of the primary poloidal field.

 (4) At the end of the previous stage of the magnetic activity before the disappearance
 of the toroidal field the poloidal field appears thanks to the $\alpha \Omega$- dynamo.

The direction of this poloidal field is close to the area of primary
poloidal field; if the latter does not experienced significant
changes. In this case, the primary poloidal field will increase.
From this poloidal field will be generated a toroidal field, which
will be stronger than the previous toroidal field. Its evolution in
a turbulent layer of the convective shell will lead to a local
activity in the medium and equatorial helio-latitudes. This activity
will be stronger than in the previous cycle. Therefore, the
"Gnevyshev-Ol's rule"  comes from the contribution of the primary
poloidal field in the evolution of the toroidal fields.

If the primary poloidal field has experienced a significant turn,
the "Gnevyshev-Ol's rule"  would be broken, as it was for 22 and 23
of the "11-year" cycles.

The rotation of the primary poloidal field can lead to the emergence
of secondary maxima of magnetic activity. These maxima are connected
with the processes of the reconnection of the lines of force of a
primary poloidal field, caught in the middle latitudes, and the
lines of force of a
 toroidal field (Fig. 6).
\begin{figure}[h!]
\centerline{
\includegraphics{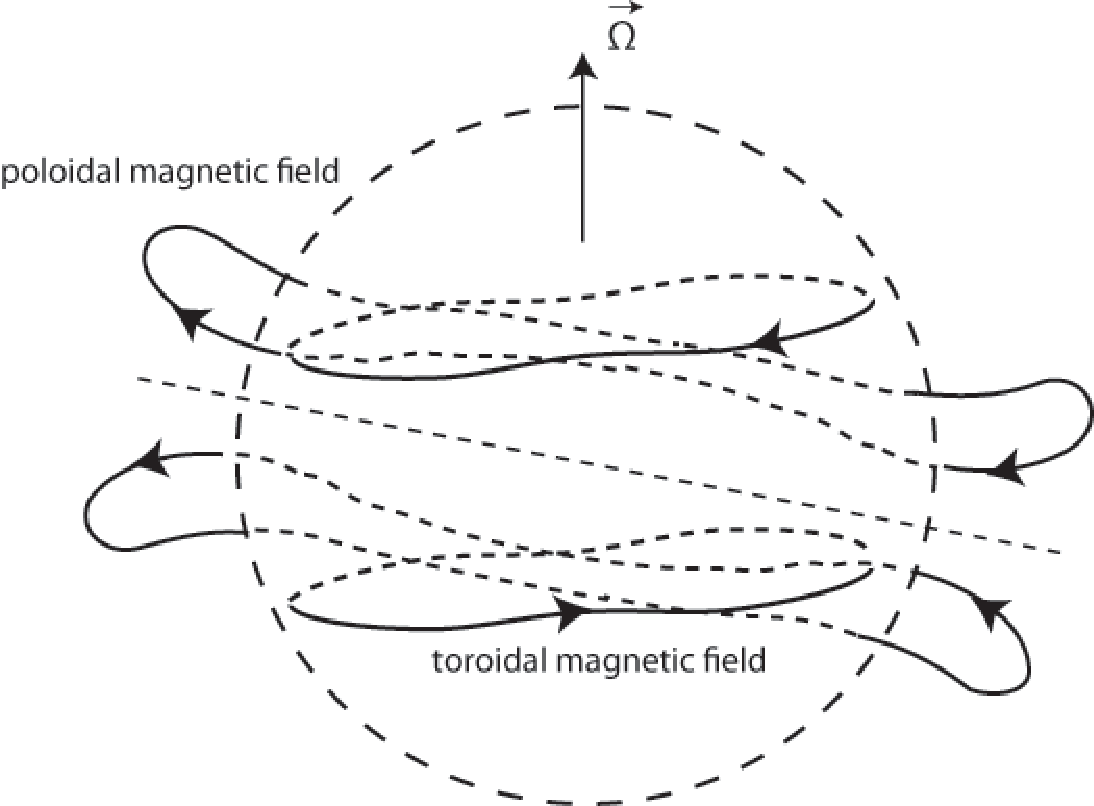}}
\caption{The intersection of the toroidal field and the turned
primary poloidal field}\label{Fi:Fig6}
\end{figure}

(5)  Energy of the global magnetic activity is determined by the
energy of the Rossby waves. Therefore, the Rossby waves are
gradually fade in the course of evolution of the primary poloidal
field. Poloidal fields can also disappear because of the
high-latitude activity. The beginning of a new cycle will move away
up to moment of formation of new Rossby waves. This may explain the
different durations of the "11-year" cycles.

(6) The existence of Rossby waves at different helio-latitudes may
lead to the existence of cycles of different duration, because the
characteristic time of these waves and poloidal fields formation
depends on helio-latitudes (the Coriolis acceleration is depended on
the latitudes)

Energy of giant convective motions at the lower laminar layer is
spent on:

- heat transfer in turbulent layer convection,

- the longitudinal transfer of momentum to maintain of the
differential rotation of the Sun,

- generation of Rossby waves.

The observed magnetic activity of the Sun has the quasistationary
character, therefore, between these branching in the process of  the
distribution of  energy of the convection, apparently, remains a
dynamic equilibrium. If, for example the formation of Rossby waves
will slow down in the mid-latitudes due to the penetration of
turbulent motions to the lower laminar layer, it will lead to the
acceleration of the meridional transport momentum that will speed up
the rotation of the mid-latitudes. The increase of the angular
velocity of rotation will lead to the reduction of the time of
Rossby waves formation (11). As a result, they will resume after a
short time.

Quantitative value of the characteristic time of Rossby waves
formation (11) depends on the thickness of the layer laminar
convection and plasma viscosity in it. We estimate the value of
viscosity, with the help of formula (11), if the time of the of
waves Rossby formation is approximately equal the duration of the
activity cycle. Taking into account that

$$ h \approx \lambda \approx \frac {l}{m}2
\pi R_I \cos \varphi \approx \frac{1,4\pi}{m} R_{Sun} \cos \varphi
$$

find the coefficient of viscosity:

$$ \nu \approx \frac {0,49 \pi^2 R_{Sun}^2 \cos^2\varphi}{m \Omega_{Sun}t_{g}^2 \sin \varphi}
\approx 1,6 \cdot 10^{12} \left( \frac{m}{6} \right)^{-1} \left(
\frac{ t_{g}}{1 year} \right)^{-2} \frac {\cos^2 \varphi}{\sin
\varphi } \frac{sm^2}{sec} \eqno (12) $$

For "active latitudes" the value of the viscosity (11) can be
compared with the value of the coefficient of viscosity
$\nu_{\gamma}$ around the base of convective envelope ($c$ is speed
of light):

$$\nu_{\gamma} \approx \frac {1}{3}\frac {c}{n \sigma_{T} } \approx 3 \cdot 10^{12} \frac {cm^2}{sec}  \eqno (13) $$

It is a coincidence speaks in favor of the above-described the
physical nature of the global magnetic activity. Approximately the
same value of turbulent viscosity is used in the $\alpha \Omega $-
dynamo model.

The model of multi-layered laminar convection using viscosity of
radiant plasma (13) is described in (Rozgacheva \& Bruevich 2002).

Rossby waves at different solar latitudes and with different
wavelengths generate a very complex structure and evolution of the
poloidal field. So check the physical picture of this poloidal field
formation according to the observations of the Sun only is
difficult.

Come to the aid of the study of solar-type stars. Using formula (11)
to assess the relationship between the duration of activity cycle
and the effective temperature of the stars.

Consider, first, that the angular velocity of rotation of the star
connected with star's effective temperature as ratio $\Omega \sim
T_{eff}~^4$, which follows from the formula (10).

Secondly, we take into account that for fully ionized hydrogen
plasma its concentration depends on the temperature, as $n \approx
T^ {\frac{3}{2}}$. Then the viscosity depends on the temperature,
see (13) as $\nu \approx \nu_\gamma \sim \frac{3}{2}$. In this case,
we can use the formula (11), and so we obtain the following
connection between the star's duration of the activity cycle and its
effective temperature:

$$ T_{cyc} \approx t_R \sim T_{eff}~^{-\frac{5}{4}}  \eqno (14) $$

The ratio (14) is consistent with the power relationships (4) and
(6) within the limits of errors. This speaks in favour of the
above-described the physical nature of the global magnetic activity
of the Sun.

For the quantitative analysis of the proposed physical picture it is
necessary to conduct the numerical experiments of the magnetic
hydrodynamics of a multilayer convective shell of the Sun. It is the
purpose of one of the following of our work.

\vskip12pt {\bf5 Conclusion}
\vskip12pt

In this paper, the following results were obtained:

 1. It is shown that the dependence of "the rotation period - the effective temperature"
 (2) is a natural law for Main sequence stars, for which
 Stefan - Boltzmann law (7), the ratio of "mass - luminosity"
 (8) and conservation of angular momentum (9) are applicable.

 2. We offer the physical picture of the interrelationship of observed properties of the
 local and global magnetic activity of the Sun.
 The main new element in this picture is a hypothesis
 about the possibility of the existence of several layers of
 the convective shell of the solar-type stars. T
 here must be at least two layers. Around the ground level the
 convective shell is the laminar convection layer, which consists of a giant convective cells.
 Thanks to the rotation on the surface of this layer Rossby waves are formed.

 These waves have spiral structure due to differential rotation.
 The formation of the laminar convection is caused by a strong heating of plasma
 photons from the zone of radiant heat transfer and high radiant viscosity of
 fully ionized dense plasma. The radiant viscosity of plasma effectively inhibits
 the directed motion of electrons more than directed motion of protons.
 Therefore, in the Rossby waves the current appears, and this current
 generates the poloidal field. This field is a primary reason of the
 whole magnetic activity of stars. Above the layer of the laminar
 convection the layer of turbulent convection extends.
 In the layer of turbulent convection the primary poloidal
 field generates a toroidal field of starts and process of generation
 of a strong local magnetic activity solar-type stars in the medium
 and equatorial latitudes starts. Local and global magnetic activity
 are interconnected thanks to the existence of internal Rossby waves and the primary poloidal field.

 3. It is shown that in the framework of the proposed hypothesis about the
 existence of internal Rossby waves you can explain the dependence
 of "the duration of the activity cycles - the effective temperature" for 11-year
 and quasi-biennial cycles, see (4) and (6). The duration of the activity cycles
 according to the order of magnitude is equal to the characteristic time of generation of Rossby waves.

 These results refer to the main problems of hydrodynamics of the Sun - differential rotation and solar dynamo.
 They point to the fact that, apparently, the model of a one-layer of turbulent
 convection and the model of turbulent theory of generation of the magnetic field
 and the magnetic activity are characteristic of young rapidly rotating stars.
 These stars may intensively lose the substance due to the strong magnetic activity.
 Their outer layers are not yet fully formed.

 In the later stages of the evolution the inner layer laminar convection is formed, because,
 firstly, the speed of rotation is reduced, and, secondly, and, secondly,
 a strong turbulent diffusion of plasma fluxes reduces the heterogeneity
 of rotation and gradually the convection passes from turbulent state to
 the laminar state. Due to this layer the phenomena of magnetic activity
 acquire the properties of orderliness, observed on the Sun.

\bigskip
\bigskip
{\bf References}

\bigskip

1. Baliunas, S.L., Donahue, R.A., et al. (1995) Astrophys. J., {\bf
438}, 269.

2. Berdugina S. V., Moss D., Sokoloff D. D., Usoskin I. G. (2006)
Astron. \& Astrophys., {\bf445}, 703.

3. Bissengaliev, R.A., Esina, Ya.V., Kuzmin, N.M., Mustsevoy, V.V.,
Khrapov S.S. (2010) Astrophysical Bulletin, {\bf66}, N3, 270.

4. Bruevich E.A. (1995) Astronomy Reports,
 {\bf{39}}, N1, 78.

5. Bruevich P.V., Bruevich E.A. (2004) Astronomical and
Astrophysical Transactions, {\bf{23}}, Issue 2, p. 165.

6. Bruevich E.A., Alekseev I.Yu. (2007) Astrophysics, {\bf{50}},
187.

7. Bruevich, E.A., Kononovich E.V. (2011) Moscow University Physics
Bull., {\bf66}, N1, 72.

8. Bruevich, E.A., Ivanov-Kholodnyj G.S. (2011) ArXiv e-prints,
    (arXiv:1108.5432v1).

9. Bruevich E.A., Rozgacheva I.K. (2010) ArXiv e-prints,
(arXiv:1012.3693v1).

10. Bruevich E.A., Rozgacheva I.K. (2012) ArXiv e-prints,
(arXiv:1204.1148v1).

11. Gilman, P.A. (1974) Ann. Rev. Astron. Astrophys., {\bf12}, 47.

12. Khramova, M.N., Kononovich, E.V. \& Krasotkin, S.A. (2002)
Astron. Vestn., {\bf 36}, 548.

13. Kichadinov, L.L. (2005) Physics-Uspekhi Journal (UFN), {\bf175},
N5, 475.

14. Kollath, Z., Olah, K. (2009) Astron. Astrophys.,  {\bf 501},
695.

15. Livshits, I.M., Obrydko, V.N. (2006) Astron. Rep., {\bf83}, N11,
1031.

16. Lockwood, G.W., Skif, B.A., Radick R.R., Baliunas, S.L.,
Donahue, R.A. and Soon W. (2007) Astrophysical Journal Suppl., {\bf
171}, 260.

17. Makarov, V.I., Ruzmaikin, A.A. and Starchenko, S.V. (1987) {\bf
111}, 267.

18. Makarov V.I.,. Obrydko V.N., Tlatov, A.G. (2001) Astron. Rep.,
{\bf78}, N9, 859.

19. Monin, A.S., (1980) Physics-Uspekhi Journal (UFN) {\bf 132}, B,
123.

20. Nordlund, A., Spruit, H.C., Ludwig, H.-G. (1997) Astron.
Astrophys., {\bf328}, 229.

21. Obrydko, V.N. (1985) Sun spots and complexes activity, Moscow,
Nauka.

22. Obrydko, V.N. (1999) Bulletin of the Russian Academy of
Sciences: Physics, {\bf63}, N11.

23. Obridko V. N., Shelting B. D., Livshits I. M., Asgarov A. B.
(2009) Solar Physics, {\bf260}, N1, 191.

24. Parker E.N., (1979) Cosmical Magnetic Fields, Clarendon Press,
Oxford.

25. Rozgacheva I.K., Bruevich E.A. (2002) Astronomical and
Astrophysical Transactions, {\bf{21}}, Issue 1, p. 27.

26. Radick, R.R., Lockwood, G.W., Skiff, B.A., Baliunas, S.L. (1998)
Astrophys. J. Suppl. Ser., {\bf 118}, 239.

27. Rivin,~Yu. R. (1989) The cycles of The Earth and the Sun,
Moscow, Nauka.

28. Seehafer, N., Gellertl, M., Kuzanyan, K.M., Pipin, V.V. (2003)
Adv. Space Res., {bf32} N10, 1819.

29. Schnerr R.S., Spruit H.C. (2010) ArXiv e-prints,
(arXiv:1010.4792v3).

30. Schmitt, D., (1987) Astron. Astrophys., {\bf174}, 281.

31. Solar and Heliospheric Observatory (1995-2012)
http://sohowww.nascom.nasa.gov/.

32. Somov, B.V. (2006) Plasma Astrophysics. Part II, Reconnection
and Flares, Springer, New York.

33. Spruit, H.C. (1998) Solar irradiance variations: theory.
Proc.IAU Symposium, 103.

34. Spruit, H.C. (2004) ArXiv e-prints, (arXiv:1004.4545v1).

35. Vainshtein, S.I., Zeldovich, YA.B. , Ruzmaikin, A.A. (1980)
Turbulent Dynamo in astrophysics. Moscow, Nauka. 29.

36. Vitinsky,~Yu.I., Kopecky,~M., Kuklin,~G.B. (1986) The statistics
of the spot generating activity of the Sun, Moscow, Nauka.

\end{document}